\documentstyle[twoside,epsf]{article}

\catcode`\@=11
\long\def\@makefntext#1{
\protect\noindent \hbox to 3.2pt {\hskip-.9pt  
$^{\mbox{\eightrm\@thefnmark}}$\hfil}\,#1\hfil}		

\def\thefootnote{\fnsymbol{footnote}}
\def\@makefnmark{\hbox to 0pt{$^{\mbox{\footnotesize \@thefnmark}}$\hss}} 
	
\def\ps@myheadings{\let\@mkboth\@gobbletwo
\def\@oddhead{\hbox{}
\rightmark\hfil\eightrm\thepage}   
\def\@oddfoot{}\def\@evenhead{\eightrm\thepage\hfil
\leftmark\hbox{}}\def\@evenfoot{}
\def\sectionmark##1{}\def\subsectionmark##1{}}

\oddsidemargin=\evensidemargin
\renewcommand{\thefootnote}{\fnsymbol{footnote}}

\newcounter{sectionc}
\newcounter{subsectionc}
\newcounter{subsubsectionc}

\renewcommand{\section}[1] {\vspace{12pt}
        \refstepcounter{sectionc} 
	\setcounter{subsectionc}{0}\setcounter{subsubsectionc}{0}\noindent 
	{\tenbf\thesectionc. #1}\par\vspace{5pt}}

\renewcommand{\subsection}[1] {\vspace{12pt}
        \refstepcounter{subsectionc} 
	\setcounter{subsubsectionc}{0}\noindent 
	{\bf\thesectionc.\thesubsectionc. {\kern1pt \bfit #1}}
	\par\vspace{5pt}}

\renewcommand{\subsubsection}[1] {\vspace{12pt}\addtocounter{subsubsectionc}{1}
	\noindent{\tenrm\thesectionc.\thesubsectionc.\thesubsubsectionc.
	{\kern1pt \tenit #1}}\par\vspace{5pt}}

\newcommand{\nonumsection}[1] {\vspace{12pt}\noindent{\tenbf #1}
	\par\vspace{5pt}}

\newcounter{appendixc}
\newcounter{subappendixc}[appendixc]
\newcounter{subsubappendixc}[subappendixc]

\renewcommand{\theappendixc}{\Alph{appendixc}}	
\renewcommand{\thesubappendixc}{\Alph{appendixc}.\arabic{subappendixc}}
\renewcommand{\thesubsubappendixc}
	{\Alph{appendixc}.\arabic{subappendixc}.\arabic{subsubappendixc}}

\renewcommand{\appendix}[1] {\vspace{12pt}
        \refstepcounter{appendixc}
	\setcounter{subsectionc}{0}	
	\setcounter{subsubsectionc}{0}	
        \setcounter{figure}{0}
        \setcounter{table}{0}
        \setcounter{lemma}{0}
        \setcounter{theorem}{0}
        \setcounter{corollary}{0}
        \setcounter{definition}{0}
        \setcounter{equation}{0}
        \renewcommand{\thefigure}{\Alph{appendixc}.\arabic{figure}}
        \renewcommand{\thetable}{\Alph{appendixc}.\arabic{table}}
        \renewcommand{\theappendixc}{\Alph{appendixc}}
        \renewcommand{\thelemma}{\Alph{appendixc}.\arabic{lemma}}
        \renewcommand{\thetheorem}{\Alph{appendixc}.\arabic{theorem}}
        \renewcommand{\thedefinition}{\Alph{appendixc}.\arabic{definition}}
        \renewcommand{\thecorollary}{\Alph{appendixc}.\arabic{corollary}}
        \renewcommand{\theequation}{\Alph{appendixc}.\arabic{equation}}
        \noindent{\tenbf Appendix \theappendixc.~~#1}\par\vspace{5pt}}

\newcommand{\subappendix}[1] {\vspace{12pt}
        \refstepcounter{subappendixc}
        \noindent{\bf Appendix \thesubappendixc. {\kern1pt \bfit #1}}
	\par\vspace{5pt}}
\newcommand{\subsubappendix}[1] {\vspace{12pt}
        \refstepcounter{subsubappendixc}
        \noindent{\rm Appendix \thesubsubappendixc. {\kern1pt \tenit #1}}
	\par\vspace{5pt}}

\newcommand{\textlineskip}{\baselineskip=13pt}
\newcommand{\smalllineskip}{\baselineskip=10pt}

\newcommand{\copyrightheading}[1]
	{\vspace*{-2.5cm}\smalllineskip{\flushleft
	{\footnotesize Quantum Information and Computation, 
	Vol.~1, No.~0 (2001) 000--000 #1}\\
	{\footnotesize \copyright\kern2pt Rinton Press}\\
	 }}

\newcommand{\publisher}[2]{{\begin{center}\footnotesize\smalllineskip 
	Received #1\\
	Revised #2
	\end{center}
	}}

\def\abstracts#1#2#3{{
	\centering{\begin{minipage}{4.5in}\footnotesize\baselineskip=10pt
	\parindent=0pt #1\par 
	\parindent=15pt #2\par
	\parindent=15pt #3
	\end{minipage}}\par}} 

\def\keywords#1{{
	\centering{\begin{minipage}{4.5in}\footnotesize\baselineskip=10pt
	{\footnotesize\it Keywords}\/: #1
	 \end{minipage}}\par}}

\renewenvironment{thebibliography}[1]
        {\frenchspacing
	 \ninerm\baselineskip=11pt
         \begin{list}{\arabic{enumi}.}
        {\usecounter{enumi}\setlength{\parsep}{0pt}     
	 \setlength{\leftmargin 12.7pt}{\rightmargin 0pt}
         \setlength{\itemsep}{0pt} \settowidth
	{\labelwidth}{#1.}\sloppy}}{\end{list}}

\newcounter{itemlistc}
\newcounter{romanlistc}
\newcounter{alphlistc}
\newcounter{arabiclistc}
\newenvironment{itemlist}
    	{\setcounter{itemlistc}{0}
	 \begin{list}{$\bullet$}
	{\usecounter{itemlistc}
	 \setlength{\parsep}{0pt}
	 \setlength{\itemsep}{0pt}}}{\end{list}}

\newcommand{\fcaption}[1]{
        \refstepcounter{figure}
        \setbox\@tempboxa = \hbox{\footnotesize Fig.~\thefigure. #1}
        \ifdim \wd\@tempboxa > 5in
           {\begin{center}
        \parbox{5in}{\footnotesize\smalllineskip Fig.~\thefigure. #1}
            \end{center}}
        \else
             {\begin{center}
             {\footnotesize Fig.~\thefigure. #1}
              \end{center}}
        \fi}

\newcommand{\tcaption}[1]{
        \refstepcounter{table}
        \setbox\@tempboxa = \hbox{\footnotesize Table~\thetable. #1}
        \ifdim \wd\@tempboxa > 5in
           {\begin{center}
        \parbox{5in}{\footnotesize\smalllineskip Table~\thetable. #1}
            \end{center}}
        \else
             {\begin{center}
             {\footnotesize Table~\thetable. #1}
              \end{center}}
        \fi}

\def\@citex[#1]#2{\if@filesw\immediate\write\@auxout
	{\string\citation{#2}}\fi
	\def\@citea{}\@cite{\@for\@citeb:=#2\do
	{\@citea\def\@citea{,}\@ifundefined
	{b@\@citeb}{{\bf ?}\@warning
	{Citation `\@citeb' on page \thepage \space undefined}}
	{\csname b@\@citeb\endcsname}}}{#1}}

\newif\if@cghi

\def\cite{\@ifnextchar[{\@tempswatrue\@citex}{\@tempswafalse\@citex[]}}

\def\citelow{\@cghifalse\@ifnextchar [{\@tempswatrue
	\@citex}{\@tempswafalse\@citex[]}}
\def\@cite#1#2{{$\null^{#1}$\if@tempswa\typeout
	{IJCGA warning: optional citation argument 
	ignored: `#2'} \fi}}

\def\@ncitex[#1]#2{\if@filesw\immediate\write\@auxout
	{\string\citation{#2}}\fi
	\def\@citea{}\@ncite{\@for\@citeb:=#2\do
	{\@citea\def\@citea{,}\@ifundefined
	{b@\@citeb}{{\bf ?}\@warning
	{Citation `\@citeb' on page \thepage \space undefined}}
	{\csname b@\@citeb\endcsname}}}{#1}}

\newif\if@cghi

\def\ncite{\@ifnextchar[{\@tempswatrue\@citex}{\@tempswafalse\@ncitex[]}}

\def\@ncite#1#2{{$#1$\if@tempswa\typeout
	{IJCGA warning: optional citation argument 
	ignored: `#2'} \fi}}

\def\pmb#1{\setbox0=\hbox{#1}
	\kern-.025em\copy0\kern-\wd0
	\kern.05em\copy0\kern-\wd0
	\kern-.025em\raise.0433em\box0}

\def\fnt#1#2{\footnotetext{\kern-.3em
	{$^{\mbox{\scriptsize #1}}$}{#2}}}

\def\fpage#1{\begingroup
\voffset=.3in
\thispagestyle{empty}\begin{table}[b]\centerline{\footnotesize #1}
	\end{table}\endgroup}

\def\runninghead#1#2{\pagestyle{myheadings}
\markboth{{\protect\footnotesize\it{\quad #1}}\hfill}
{\hfill{\protect\footnotesize\it{#2\quad}}}}
\font\tenrm=cmr10
\font\tenit=cmti10 
\font\tenbf=cmbx10
\font\bfit=cmbxti10 at 10pt
\font\ninerm=cmr9

\font\eightrm=cmr8

\def\FigName{figure}%
\newbox\captionbox
\long\def\@makecaption#1#2{%
  \ifx\FigName\@captype
    \vskip\abovecaptionskip
    \setbox\tempbox\hbox{{\figurecaptionfont #1\hskip1em #2}}
	\ifdim\wd\tempbox< 28pc
	\centerline{\box\tempbox}
	\else
	{\figurecaptionfont #1\hskip1em #2\par}
\fi\else
  	\setbox\tempbox\hbox{{\tablecaptionfont #1\hskip1em #2}}
 	\ifdim\wd\tempbox< 28pc 
	\centerline{\box\tempbox}
	\else
	{\tablecaptionfont #1\hskip1em #2\par}%
	\fi   
 \vskip\belowcaptionskip
 \fi}
\InputIfFileExists{psfig.sty}{\typeout{psfig.sty inputed...ok}}
			{\typeout{Warning: psfig.sty could be be found.}}
\InputIfFileExists{epsfsafe.tex}{\typeout{epsfsafe.tex inputed...ok}}
			{\typeout{Warning: epsfsafe.tex could not be found.}}
\InputIfFileExists{epsfig.sty}{\typeout{epsfig.sty inputed...ok}}
		{\typeout{Warning: epsfig.sty could not be found.}}
\InputIfFileExists{epsf.sty}{\typeout{epsf.sty inputed...ok}}
	{\typeout{Warning: epsf.sty could not be found.}}
\def\fps@figure{tbp}
\def\ftype@figure{1}
\def\ext@figure{lof}
\def\fnum@figure{Fig.\ \thefigure}
\def\qed{\hbox{${\vcenter{\vbox{	          
   \hrule height 0.4pt\hbox{\vrule width 0.4pt height 6pt
   \kern5pt\vrule width 0.4pt}\hrule height 0.4pt}}}$}}

\renewcommand{\thefootnote}{\fnsymbol{footnote}}  

\def\<{\langle}
\def\>{\rangle}
\def\be{\begin{equation}}
\def\ee{\end{equation}}
\def\bea{\begin{eqnarray}}
\def\eea{\end{eqnarray}}

\def\lbm{ \left[\rule{0pt}{2.1ex}\right. }
\def\rbm{ \left.\rule{0pt}{2.1ex}\right] }

\begin{document}

\setlength{\textheight}{7.7truein}    

\runninghead{Quantum Vernam Cipher} 
            {Debbie W. Leung}

\normalsize\textlineskip
\thispagestyle{empty}
\setcounter{page}{1}

\copyrightheading{}	

\vspace*{0.88truein}

\fpage{1}


\centerline{\bf QUANTUM VERNAM CIPHER}

\vspace*{0.37truein}


\centerline{\footnotesize DEBBIE W. LEUNG}
\vspace*{0.015truein}

\centerline{\footnotesize\it IBM T.J. Watson Research Center, P.O. Box 218}
\baselineskip=10pt
\centerline{\footnotesize\it Yorktown Heights, New York 10598, USA}

\vspace*{0.225truein}

\publisher{(received date)}{(revised date)}

\vspace*{0.21truein}

%
\abstracts{
We discuss aspects of secure quantum communication by proposing and
analyzing a quantum analog of the Vernam cipher (one-time-pad).
The quantum Vernam cipher uses entanglement as the key to encrypt 
quantum information sent through an insecure quantum channel.
First, in sharp contrast with the classical Vernam cipher, the
quantum key can be recycled securely.
We show that key recycling is intrinsic to the quantum cipher-text, 
rather than using entanglement as the key.
Second, the scheme detects and corrects for arbitrary transmission
errors, and it does so using only local operations and classical
communication (LOCC) between the sender and the receiver.
The application to quantum message authentication is discussed. 
Quantum secret sharing schemes with similar properties are 
characterized. 
We also discuss two general issues, the relation between secret
communication and secret sharing, the classification of secure
communication protocols.}{}{}

\vspace*{10pt}





\setcounter{footnote}{0}
\renewcommand{\thefootnote}{\alph{footnote}}

\vspace*{1pt}\textlineskip

\section{Introduction}
\label{sec:intro}

\noindent
Recent developments in quantum information theory have brought many
surprises in cryptology.  A partial list includes an efficient
quantum algorithm for factoring~\cite{Shor94} which can break the
condition for security in many cryptographic protocols,
unconditionally secure quantum key distribution
protocols~\cite{BB84,Ekert91,Mayers96,Lo99b}, and a no-go theorem for
unconditionally secure quantum bit
commitment~\cite{Mayers97bc,Lo97bc}.
Cryptographic protocols for quantum information are also being developed.
For examples, see
Refs.~\ncite{Mosca00,Boykin00,Cleve99,Gottesman99b,Barnum01,Crepeau01}.

Emerging from these interesting results are important open questions
on what quantum mechanics admits and prohibits in cryptography and the
reasons why.
This paper reports partial progress along this direction, by analyzing
a proposed ``quantum Vernam cipher'' which encrypts a quantum plain-text
to a quantum cipher-text using {\em entanglement} as a ``key''.
The proposed scheme is a quantum analog of various existing schemes,
including the classical Vernam cipher~\cite{Vernam26} (one-time-pad)
in which all of the plain-text, the cipher-text, and the key are
classical, the eavesdrop-detecting channel~\cite{BBB82}, in which the
plain-text and the key are classical but the cipher-text is quantum,
and the private quantum channel~\cite{Mosca00,Boykin00} in which the
plain-text and the cipher-text are quantum but the key is classical.

One intriguing property of the quantum Vernam cipher is that the key
can be recycled securely using test and purification procedures for
entanglement~\cite{Lo99b,Bennett96a}.\footnote{
A recent article~\cite{Zhang00} has {\em independently} reported using
entanglement as a recyclable quantum key to conceal classical
information.  The application to encrypt quantum information was
suggested but not accomplished~\cite{Note}.}~~
As a comparison, key recycling is insecure in the classical Vernam
cipher~\cite{Shannon49b} but secure in the eavesdrop-detecting
channel~\cite{BBB82}.
These observations suggest that the security of key recycling comes
{from} the possibility to detect eavesdropping in the quantum
cipher-text, rather than using entanglement as a key.  We give
further support to this suggestion by modifying the private quantum
channel to securely recycle the classical key.

Another intriguing property of the quantum Vernam cipher is the
ability to correct for {\em any} damage on the transmitted quantum
state.
Moreover, the correction procedure involves only classical
communication between the sender and the receiver.
These can be explained by the theory of quantum secret
sharing~\cite{Cleve99,Gottesman99b}.  Quantum secret sharing schemes
with similar properties are characterized.
We discuss general connections between secret communication and secret
sharing, and apply the connections to other secret communication
schemes.

As suggested by the above results, and in concert with our effort to
relate cryptographic properties to various elements in cryptographic
schemes, we classify existing schemes according to the classical
or quantum nature of the communication channel and the key (the
resources) and the plain-text (the application), and consider 
the security of key recycling and reliability for each class.
Besides the schemes mentioned above, teleportation~\cite{Bennett93},
superdense coding~\cite{Bennett92}, and key distribution
protocols~\cite{BB84,Ekert91} are also included.

Secure key recycling and reliability are closely related to message
authentication.  We briefly discuss applications of our analysis to
the authentication of quantum messages~\cite{Barnum01,Crepeau01}.

Despite the fact that entanglement is recycled in the quantum Vernam
cipher, we find that, given the same resources, secure quantum
communication can be more efficiently realized by distributing 
entanglement and then teleporting the state.
We emphasize that our main goal is to understand and analyze security
in quantum protocols; our proposed cipher and the comparisons with 
other schemes are tools for doing so.  

This paper is structured as follows.  The quantum Vernam cipher is
described in Section~\ref{sec:all1tp} following the reviews of private
key encryption and the private quantum channel.  Eavesdropping and
error correction strategies are explained in Section~\ref{sec:eec}.
Key recycling is analyzed in Section~\ref{sec:recycle}.  The
connections between secret communication and secret sharing are
discussed in Section~\ref{sec:ss}.  We conclude with a classification
of secure communication protocols, some applications of the analysis
to authentication, and some open questions.  For completeness, various
relevant cryptographic schemes are described in
Appendices~\ref{sec:ciphers},~\ref{sec:polycode},
and~\ref{sec:teleport}.

\subsection{Definitions and Assumptions}

In communication problems, the sender, the receiver, and any adversary
(such as an eavesdropper) are traditionally called Alice, Bob, and Eve
respectively.
For simplicity in notation and in the proofs, we make 
the following assumptions throughout the paper.
\begin{itemlist}
\item Channel noise and logical errors are negligible. 
\item Alice and Bob have a 2-way classical {\em broadcast} channel.
Hence classical communication is public but unjammable and
authenticated (not forged or tampered with).
\item Alice and Bob may also be given a quantum channel or entanglement.
Such quantum channel is assumed insecure, while the given entanglement
is pure and authenticated.
\end{itemlist}
The two quantum resources are inequivalent.  
Entanglement can be converted to a secure quantum channel 
by teleportation (see Appendix \ref{sec:ciphers}).
A quantum channel which is insecure can establish ``mixed
entanglement'', but further test and distillation
procedures~\cite{Bennett96a,Lo99b} are needed to establish pure
entanglement.

\section{Concealing Ciphers}
\label{sec:all1tp}

\noindent
In this section, we describe the quantum Vernam cipher.  We first
review basic notions in private key encryption, using the classical
Vernam cipher, the eavesdrop-detecting channel, and the private
quantum channel as examples.  These examples also motivate the
construction of the quantum Vernam cipher.
We concentrate on the ability to conceal the communicated secret from
an eavesdropping adversary.  Other aspects of security will be
discussed later.

\subsection{Private key encryption} 
\label{subsec:c1tp}

\noindent
%
%
In secret classical communication using private key encryption, Alice
and Bob share a secret string $K$, called the ``key'', which encrypts
(locks) a message $M$ from Eve during transmission and decrypts
(opens) $M$ for Bob afterwards.

For example, in the Vernam cipher~\cite{Vernam26}, a {\em random}
$n$-bit key $K$ is used to encrypt an $n$-bit message $M$ (also 
known as the {\em plain-text}).  
Alice sends a {\em cipher-text} $C = M \oplus K$ to Bob, where
$\oplus$ denotes bitwise {\sc xor} (addition modulo $2$).  
Bob decodes by calculating $C \oplus K = M$.
Shannon proved that~\cite{Shannon49b} the Vernam cipher is {\em absolutely
secure}:\footnote{
A cipher is absolutely secure if $C$ and $M$ are independent.}~
$C$ is random and {\em independent} of $M$ when $K$ is random and unknown. 
Shannon also proved that absolute security requires the entropy (thus
the length) of $K$ to be at least $n$.
Thus reusing a key, even with privacy amplification~\cite{Bennett95},
compromises security when previously transmitted cipher-text might have
been tapped.

As another example, we consider a simple case of the
eavesdrop-detecting channel~\cite{BBB82}.
Let $r$ be a security parameter.  
An $n$-bit classical plain-text $M$ is encrypted with two
$(n\!+\!r)$-bit classical keys $K_1,K_2$ into a {\em quantum}
cipher-text as follows.
Alice concatenates $M$ with $r$ random subset parities of $M$ to form
$M'$.  She sends each bit of $K_1 \oplus M'$ in the basis
$\{|0\>,|1\>\}$ or $\{|+\>,|-\>\}$ depending on each bit of $K_2$.
After Bob receives and decodes the cipher-text, Alice announces the
random subsets.
The decoded message is accepted only if all the subset parities are
correct.  
The security has been analyzed for the intercept-resend attack.  When
$l$ bits of the cipher-text have been intercepted, the probability to
have no inconsistencies in the subset parities is no more than $({3
\over 4})^{l}$, and the keys can be reused with privacy 
amplification.  
The security against a more general attack, and bounds on the
information gain by Eve is not available in the literature.  A similar
analysis for a different scheme is presented in
Sec.~\ref{sec:recycle}.\ref{subsec:ckrecycle}.
%

\subsection{Private quantum channel}
\label{subsec:pqc}

\begin{figure}[ht]
\vspace*{-1ex}
\begin{center}
\mbox{\psfig{file=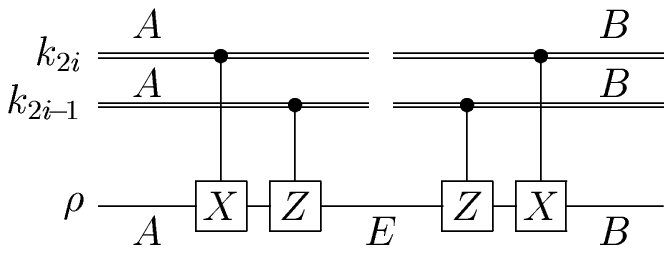,width=2.4in}}
\vspace*{1.5ex}
\fcaption{The private quantum channel.  Time runs from left to right.  
The symbols $A$, $B$, and $E$ stand for Alice, Bob, and Eve and denote
the owners of the registers.
Double lines represent classical bits.  $X$ and $Z$ are applied to the
quantum state if their respective classical control bits equal $1$.
These conventions are assumed throughout the paper. 
}
\label{fig:scheme0}
\end{center}
\end{figure}

\noindent
We motivate the quantum Vernam cipher by reviewing the following
canonical example of the private quantum
channel~\cite{Mosca00,Boykin00}, which uses a classical key to encrypt
a quantum plain-text to a quantum cipher text.
For simplicity, we call the canonical example the private quantum channel.  
Let 
\vspace*{-3ex}
\bea
	I &=& \left[ \begin{array}{cc} 1 & 0 \\ 0 & 1 \end{array} \right]\,,
~~~	Z  =  \left[ \begin{array}{cc} 1 & 0 \\ 0 & -1 \end{array} \right]\,,
\nonumber
\\
	X &=& \left[ \begin{array}{cc} 0 & 1 \\ 1 & 0 \end{array} \right]\,,
~~~	ZX = \left[ \begin{array}{cc} 0 & 1 \\ -1 & 0 \end{array} \right]\,,
\eea
denote the $2 \times 2$ identity and three ``Pauli matrices''.
To send one quantum bit (qubit) given by the density matrix $\rho$, 
Alice and Bob share a $2$-bit key $K = (k_1,k_2)$.  
Alice applies $Z^{k_2} X^{k_1}$ to $\rho$ and sends to Bob the resulting 
``cipher-text'' $\rho' = Z^{k_2} X^{k_1} \rho X^{k_1} Z^{k_2}$, 
which is decoded by Bob by applying $X^{k_1} Z^{k_2}$.
{From} Eve's point of view, Alice is sending $\rho$, $X \rho X$, $Z
\rho Z$, and $ZX \rho XZ$ at random; she sees a mixture $\frac{I}{2}$ 
which is independent of $\rho$.
To send an $n$-qubit state $\rho$, the $1$-qubit scheme is applied bitwise.
Let $K$ be a $2n$-bit classical key with $i$-th bit $k_i$.
Let $X_i$ and $Z_i$ denote $X$ and $Z$ acting on the $i$-th qubit.
Alice sends to Bob 
$\rho' = U_K \rho \hspace*{0.5ex} U_K^\dagger$, where  
$U_K = \bigotimes_i Z_i^{k_{2i}} X_i^{k_{2i-1}}$.  
Bob applies $U_K^\dagger$ to recover $\rho$ from $\rho'$.  
Eve sees a mixture of uniformly distributed possible states: 
\vspace*{-1ex}
\be
	  \frac{1}{2^{2n}} \sum_K U_K \rho \hspace*{0.5ex} U_K^\dagger 
	= \frac{1}{2^n} I^{\otimes n} 
\label{eq:randomize}
\,,
\ee
which is independent of $\rho$.\footnote{  
%
Equation~(\ref{eq:randomize}) can be derived using the Pauli
decomposition of $\rho$: each nontrivial component anticommutes with
half of the $U_K$ and vanishes in the sum, leaving only the identity
term.}~~
It was also proved in Ref.~\ncite{Mosca00} that $H(K) \geq 2n$ is
necessary to completely randomize an arbitrary $\rho$.
A schematic diagram is given in Fig.~\ref{fig:scheme0}.  

\clearpage

\subsection{The quantum Vernam cipher}
\label{subsec:q1tp}

\begin{figure}[ht]
\begin{center}
\mbox{\psfig{file=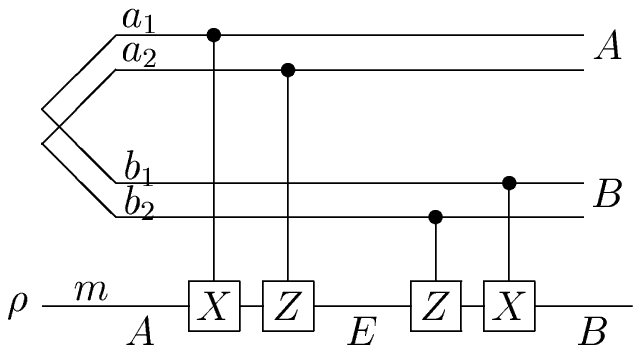,width=2.4in}}
\vspace*{2ex}
\fcaption{The quantum Vernam cipher to send one qubit.}
\label{fig:scheme1}
\end{center}
\end{figure}

\noindent
We use entanglement as the key in our quantum Vernam cipher.
The fundamental unit of entanglement is an ``ebit''.  
Alice and Bob are said to share an ebit if each possesses one qubit of
a {\em known} maximally entangled state of two qubits, such as the EPR
states $|\Phi^{\pm}\> = {1 \over \sqrt{2}}(|00\> \pm |11\>)$.
The procedure to transmit one qubit using two ebits is summarized in 
Fig.~\ref{fig:scheme1}. 
\\[1.5ex] 
The registers in Fig.~\ref{fig:scheme1} are labeled by
$a_1$, $a_2$, $b_1$, $b_2$, and $m$.
The two registers $a_1,b_1$ are initially in the state
$|\Phi^+\>$, and so are $a_2,b_2$.
The registers $a_1,a_2$ belong to Alice and $b_1,b_2$ belong to Bob
all the time.
The register $m$ initially carries the message $\rho$ and belongs to Alice. 
Alice applies a controlled-$X$ ({\sc cnot}) from $a_1$ to $m$ and 
a controlled-$Z$ ({\sc cz}) from $a_2$ to $m$ and sends $m$ to Bob. 
We assume Eve takes control of $m$ during the transmission.
When Bob receives $m$, he applies a {\sc cz} from $b_2$ to $m$,
followed by a {\sc cnot} from $b_1$ to $m$ to recover $\rho$.
To send an $n$-qubit state $\rho$, the one-qubit protocol is applied
bitwise.
We show that the quantum Vernam cipher is a {\em purification} of the
private quantum channel, superposing all possible key states:
The key registers $(a_1,b_1,\cdots,a_{2n},b_{2n})$ have initial state
$|\Phi^+\>^{\otimes 2n}$.
Reordering the qubits as $(a_1,\cdots,a_{2n},b_1,\cdots,b_{2n})$, the
initial key state ${1 \over 2^{n}} \sum_K |K\>|K\>$, where $K$ ranges
over all $2n$-bit strings, is indeed the superposition of all possible
classical keys.
Finally, the quantum Vernam cipher and the private quantum channel
have {\em equivalent} encoding and decoding operations, establishing 
the claim.
Eve sees a cipher-text described by tracing out the subsystem
$\{a_1,\cdots,a_{2n},b_1,\cdots,b_{2n}\}$, which corresponds to
averaging over all possible keys $|K\>|K\>$.
Following the discussion in Section~\ref{sec:all1tp}.\ref{subsec:pqc},
Eve sees the state $I^{\otimes n}/2^n$.
Thus Eve obtains no information on $\rho$.  

In the absence of eavesdropping, the circuit in Fig.~\ref{fig:scheme1}
acts trivially, so that $\rho$ is recovered, and the key
$|\Phi^+\>^{\otimes 2n}$ is regenerated.
We now consider the effects of eavesdropping. 

\section{Eavesdropping and Error Correction}
\label{sec:eec}

Even though Eve obtains no information from the cipher-text, she may
disturb, destroy, or alter it, and entangle her ancilla with the
quantum key to be regenerated.
In this section, we describe the effects of eavesdropping and a basic
correction method, which are starting points for our discussions in
Sections~\ref{sec:recycle} and~\ref{sec:ss}.

\subsection{General eavesdropping and correction strategy}
\label{subsec:genform}

\noindent
We assume that the plain-text is initially disentangled from Eve. 
Eve's most general strategy is to apply a joint unitary operation $U$
on the transmitted cipher-text and a pure state ancilla of hers, and
send Bob ``something''.
We may assume she outputs the correct number of qubits as Bob can
add or discard qubits.
Note that there is no further communication from Eve to Alice or Bob.
Thus subsequent action $\cal F$ by Eve on her ancilla {\em cannot}
change the superoperator ${\cal E}$ that describes the transmission of
the cipher-text.  The situation is summarized as
\bea
\centerline{\mbox{\psfig{file=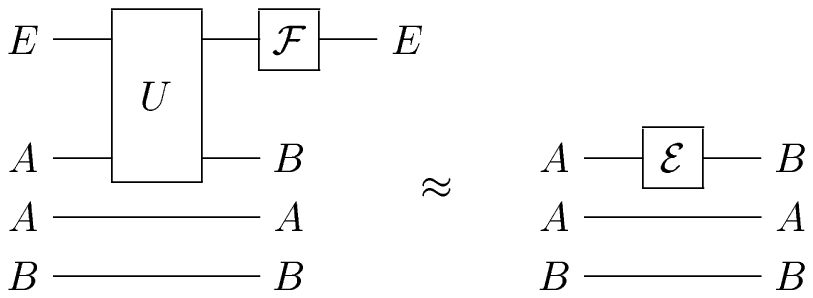,width=2.8in}}}
\nonumber
\eea
A convenient representation of $\cal E$ is given 
by~\cite{Chuang96d,Knill96b}  
\be
	{\cal E}(\rho) = \sum_{ij} e_{ij} P_i \rho P_j^\dagger
\label{eq:chirep}
\ee
where $e_{ij}$ are entries of a positive matrix and the sum is over
all Pauli matrices $P_i$ on the $n$-qubit cipher-text.
Equation (\ref{eq:chirep}) can be interpreted as a process that
transforms a state $\rho$ into a mixture $\sum_k D_k \rho D_k^\dagger$
where $D_k$ are noninterfering errors.  Expressing each $D_k$ as a
linear combination of Pauli matrices, one obtains Eq.~(\ref{eq:chirep}).
The $P_i$ in Eq.~(\ref{eq:chirep}) thus represent errors that may
interfere with each other \cite{interfere}.
Using the language of quantum error correction, we call the $P_i$ Pauli 
errors.  
We now show that if Alice and Bob determine with high probability what
Pauli error has occurred, their final state is almost disentangled from 
Eve.  The process of determining the error is called {\em syndrome
extraction}.

The cipher-text is generally {\em part of} a state $\tilde{\rho}$
obtained from encoding the plain-text with some ancilla.
The state possessed by Alice and Bob after the transmission is given by
\be
	({\cal I} \otimes {\cal E})(\tilde{\rho}) 
	= \sum_{ij} e_{ij} 
	(I \otimes P_i) \tilde{\rho} (I \otimes P_j^\dagger) 
\label{eq:transtate}
\ee
where the identity operator ${\cal I}$ acts on the uncommunicated 
subsystem.  

First suppose it is possible to {\em perfectly} distinguish the states 
$(I \otimes P_i) \tilde{\rho} (I \otimes P_i^\dagger)$ nondestructively.
Then, there is a projective measurement $\cal Q$ with projectors $Q_i$ 
such that 
\bea
	& {\rm if~} j=i & 
	Q_i (I \otimes P_j) \tilde{\rho} (I \otimes P_j^\dagger) Q_i 
	= (I \otimes P_j) \tilde{\rho} (I \otimes P_j^\dagger) 
\label{eq:yes}
\;,
\\
	& {\rm if~} j \neq i & 
	Q_i (I \otimes P_j) \tilde{\rho} (I \otimes P_j^\dagger) Q_i = 0
\label{eq:no}
\;.
\eea
Since $\tilde{\rho}$ is positive, Eqs.~(\ref{eq:yes}) and (\ref{eq:no}) 
are equivalent to 
\bea
	& {\rm if~} j=i & 
	Q_i (I \otimes P_j) \tilde{\rho} = (I \otimes P_j) \tilde{\rho}  
\label{eq:yes1}
\;, 
\\
	& {\rm if~} j \neq i & 
	Q_i (I \otimes P_j) \tilde{\rho} = 0
\label{eq:no1}
\;.
\eea
The projector $Q_i$ {\em removes} any term in Eq.~(\ref{eq:transtate})
with a $P_j$ for all $j \neq i$, leaving only the output $(I \otimes
P_i) \tilde{\rho} (I \otimes P_i^\dagger)$,
which is independent of ${\cal E}$ and disentangled from Eve.

We will consider situations deviating from the above perfect scenario.
For example, the measurement outcome $i$ may be accompanied by some
irreversible state change $O_i$.
Moreover, the measurement may only distinguish subsets of errors or be
probabilistic, so that multiple terms in Eq.~(\ref{eq:transtate}) may
remain in the final state.
In any case, if a syndrome $i$ is extracted with high probability, the
post-measurement state has density matrix dominated by $O_i (I \otimes
P_i) \tilde{\rho} (I \otimes P_i^\dagger) O_i^\dagger$, and is almost
disentangled from Eve.

Suppose Alice and Bob reuse a private key obtained from 
$({\cal I} \otimes {\cal E})(\tilde{\rho})$ which is entangled with Eve. 
Eve can learn about the future communication or correlate different
rounds of communicated materials only through the correlation with the
reused private key.  Such correlation is small when syndrome
extraction succeeds with high probability, in which case Eve has 
little information on any nontrivial function on all the plain-text.  
Key recycling is then {\em semantically secure}~\cite{GB97}.  
A scheme is {\em semantically secure}~\cite{GB97}, if there is
vanishing difference between the probabilities to estimate the 
value of any nontrivial function on the plain-text, 
with or without the cipher-text.
%


\subsection{Error correction for the quantum Vernam cipher}
\label{subsec:q1tpec}

\noindent
Recall that it suffices to identify the Pauli error that occurs in the
cipher-text.  We show how this can be done perfectly in the quantum
Vernam cipher.
We use the fact that Fig.~\ref{fig:scheme1} acts trivially, and the
commutation relations
\bea
\mbox{\psfig{file=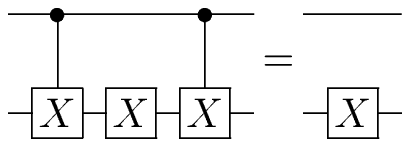,width=1.3in}}\,,~~~~~~~~
\mbox{\psfig{file=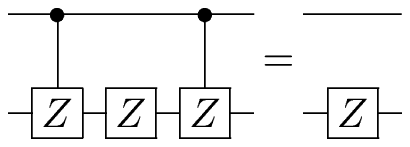,width=1.3in}}\,,
\nonumber
\\[0.2ex]
\nonumber
\\
\mbox{\psfig{file=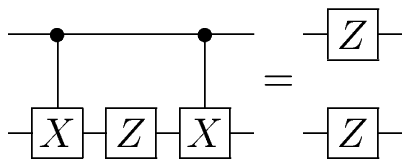,width=1.3in}}\,,~~~~~~~~
\mbox{\psfig{file=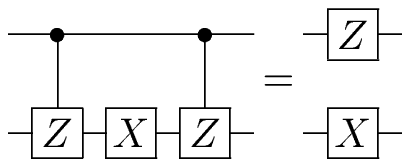,width=1.3in}}\,,
\nonumber
\vspace*{1ex}
\eea
to find the effect of errors on the cipher-text for the one-qubit
protocol:
\bea
\mbox{\psfig{file=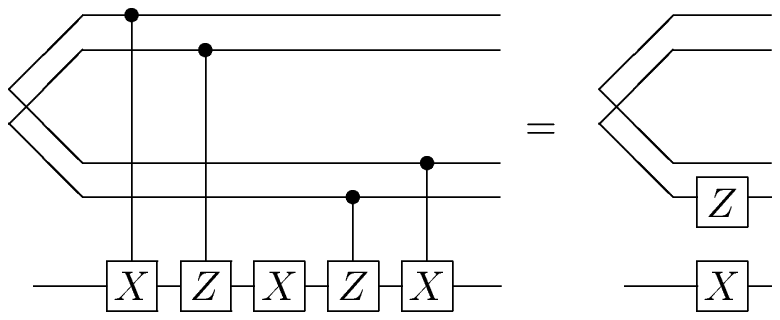,width=2.7in}}~~
\nonumber
\eea
and 
\bea
\mbox{\psfig{file=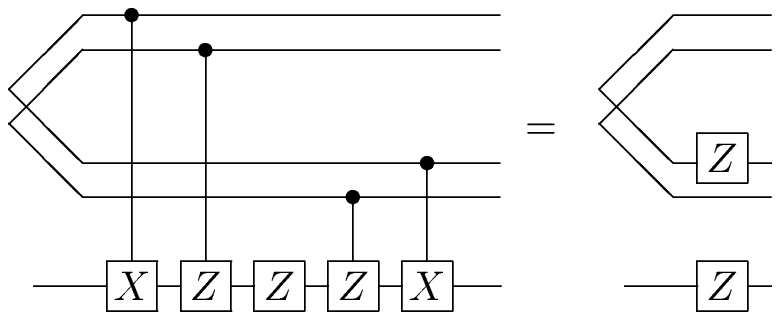,width=2.7in}}~~
\nonumber
\eea
An $X$ error in the transmitted cipher-text propagates to the decoded
message together with a $Z$ error on $b_2$, changing $\{a_2,b_2\}$
{from} $|\Phi^+\>$ into $|\Phi^-\>$.
Likewise, a $Z$ error turns $\{a_1,b_1\}$ into $|\Phi^-\>$ and an $XZ$
error turns both EPR pairs into $|\Phi^-\>$.
Alice and Bob can distinguish $|\Phi^+\>$ from $|\Phi^-\>$ by
independently measuring their halves of the EPR pair along the $|\pm\>
= {1 \over \sqrt{2}} (|0\> \pm |1\>)$ basis and comparing their
results on a broadcast channel.
Since $|\Phi^+\> = {1 \over \sqrt{2}} (|++\> + |--\>)$ and $|\Phi^-\>
= {1 \over \sqrt{2}} (|+-\> + |-+\>)$, the measured state is
$|\Phi^+\>$ ($|\Phi^-\>$) when their results agree (disagree).
Therefore, the possible Pauli errors $I$, $X$, $Z$, and $XZ$ can be
perfectly distinguished and corrected.
The same argument applies to transmitting $n$ qubits.

We emphasize that this detection procedure effectively turns Eve's
most general action into a Pauli error.
An example to recover the message without the cipher-text is
given in Appendix~\ref{sec:lostm}.

\section{Key Recycling}
\label{sec:recycle}

\noindent
We have seen that the EPR pairs in the quantum Vernam cipher can be
measured to extract the exact error syndrome.
We now show that, when many qubits are sent, it is possible to use
less entanglement (per qubit) for syndrome extraction with very 
high probability.
The remaining EPR pairs can be recycled, with semantic security.
We show strong evidence that security is due to transmitting a quantum
cipher-text, rather than using entanglement as the key, by modifying
the private quantum channel to recycle a classical key.

\subsection{Recycling quantum key}
\label{subsec:qkrecycle}

\noindent
Recall that after sending $n$ qubits with the quantum Vernam cipher,
Alice and Bob share $2n$ EPR pairs either in $|\Phi^{+}\>$ or
$|\Phi^{-}\>$, in a one-to-one correspondence with the Pauli error in
the cipher-text.
Syndrome extraction is equivalent to learning the identity of these
EPR pairs.
Asymptotically, this can be done in two steps.  
The first step, adapted from Ref.~\ncite{Lo99b}, is a preliminary test
for eavesdropping by testing if the EPR pairs are $|\Phi\>^{\otimes 2n}$.
%
%
Without indication of eavesdropping, the decoded state is accepted,
and the EPR pairs are recycled.
Otherwise, a second step is performed to find the identity of the EPR
pairs by a random hashing method adapted from Ref.~\ncite{Bennett96a}.
%
This procedure applies to the most general eavesdropping strategy.

Let the identity of the $2n$ EPR pairs be represented by a $2n$-bit
string {\bf v}, with $0$ and $1$ corresponding to $|\Phi^+\>$ and
$|\Phi^-\>$.\footnote{
%
This representation is a {\em simplified} version of that
in Ref.~\ncite{Bennett96a}.}~~
We first describe a useful protocol to obtain the {\em parity} of a
subset of bits in {\bf v}.  The ``bilateral {\sc xor}'' ({\sc bxor}),
defined as {\sc cnot}$_{a_1 a_2}$$\times${\sc cnot}$_{b_1 b_2}$,\footnote{
%
The first and second subscripts denote the control and 
target bits.}~ 
effects the transformation:
\bea
	|\Phi^+\>|\Phi^+\> \rightarrow |\Phi^+\>|\Phi^+\> &\,,~~~~&
	|\Phi^-\>|\Phi^-\> \rightarrow |\Phi^+\>|\Phi^-\> \,,
\nonumber
\\	|\Phi^+\>|\Phi^-\> \rightarrow |\Phi^-\>|\Phi^-\> &\,,~~~~& 
	|\Phi^-\>|\Phi^+\> \rightarrow |\Phi^-\>|\Phi^+\> \,,
\nonumber
\eea
where the qubits are ordered as $a_1,b_1,a_2,b_2$.  
The control pair $(a_1,b_1)$ becomes the parity of the two pairs.
Likewise, the parity of a subset $\{s_1,s_2,s_3,\cdots\}$ can be
found by applying {\sc bxor} from an extra $|\Phi^+\>$ to all of
$\{s_1,s_2,s_3,\cdots\}$.

For the preliminary test for eavesdropping, let $r$ be a security
parameter.
Alice and Bob pick $r$ random subsets of {\bf v} and find
their parities using $r$ extra $|\Phi^+\>$.\footnote{
%
These extra $|\Phi^+\>$ are unnecessary but they simplify
the procedure.}~~
If ${\bf v} = {\bf 0}$, all subsets have even parities.  Otherwise, each
random subset has equal probability to be odd or even, and the
probability of obtaining only even parities is $2^{-r}$.

If all $r$ parities are even, Alice and Bob recycle the $2n$-ebit key.
The probability for Alice and Bob to miss an error in the decoded 
message and recycle a compromised key is
\bea
	\mbox{Prob}(\mbox{pass and erroneous}) 
	\leq \mbox{Prob}(\mbox{pass}|\mbox{erroneous}) 
	= {1 \over 2^r}
\nonumber
\eea
which can be made arbitrarily small by choosing a sufficiently large $r$.  

If any subset has odd parity, Alice and Bob determine {\bf v} as follows.
The distribution of {\bf v} is generally unknown. 
However, Alice and Bob can {\em estimate} the Hamming weight~\footnote{ 
%
The Hamming weight is the number of 1s in a bit-string.}~~ 
of {\bf v} by sampling $r_2$ random bits of {\bf v}.  How $r_2$ depends 
on the security level can be found as follows.  
If the Hamming weight of {\bf v} is $\alpha n$, and $\tilde{\alpha}
r_2$ $1$'s are sampled, Chebyshev's inequality implies
$\forall \delta > 0$ $\mbox{Prob}(|\tilde{\alpha} - \alpha| \geq \delta) 
< {1 \over 4 \delta^2 r_2}$.\footnote{
%
The test bits are identically distributed, and {\em negatively} 
correlated, so that Chebyshev's inequality applies.}~~ 
Hence $\forall \epsilon > 0$, choosing $r_2 > {1 \over 4 \delta^2
\epsilon}$ guarantees $\mbox{Prob}(\alpha \in (\tilde{\alpha} -
\delta, \tilde{\alpha} + \delta)) \geq 1-\epsilon$.
%
%
Thus with probability larger than $1-\epsilon$, ${\bf v} \in \cal T$
the typical set of a binomial distribution with bias $\tilde{\alpha}$
and with size no greater than $2^{2n H(\tilde{\alpha} + \delta)}$.
Here, $H$ denotes the binary entropy function~\cite{Cover91} and
for simplicity $\tilde{\alpha} + \delta < {1 \over 2}$.
As each random subset parity eliminates about half of the possible
values of {\bf v}, {\bf v} can be identified with $r_3 \approx 2n
H(\tilde{\alpha} + \delta)$ random subset parities.
Approximately $2n (1 - H(\tilde{\alpha} + \delta))$ EPR pairs can be
recycled with vanishing correlation with Eve.

Note that the preliminary test uses $r$ ebits, and the second step uses  
$r_2 + r_3$ ebits.  
Since $r$ and $r_2$ are independent of $n$, they are negligible for
asymptotically large $n$.  In contrast, $r_3 \propto n$.  This is the
reason for splitting the procedure into two steps.
Finally, we use classical probabilities throughout the discussion
since measurements are only made in the $|\Phi^{\pm}\>$
basis~\cite{Lo99b}. \\[2ex]

\subsection{Recycling classical key}
\label{subsec:ckrecycle}

\begin{figure}[ht]
\vspace*{1ex}
\begin{center}
\mbox{\psfig{file=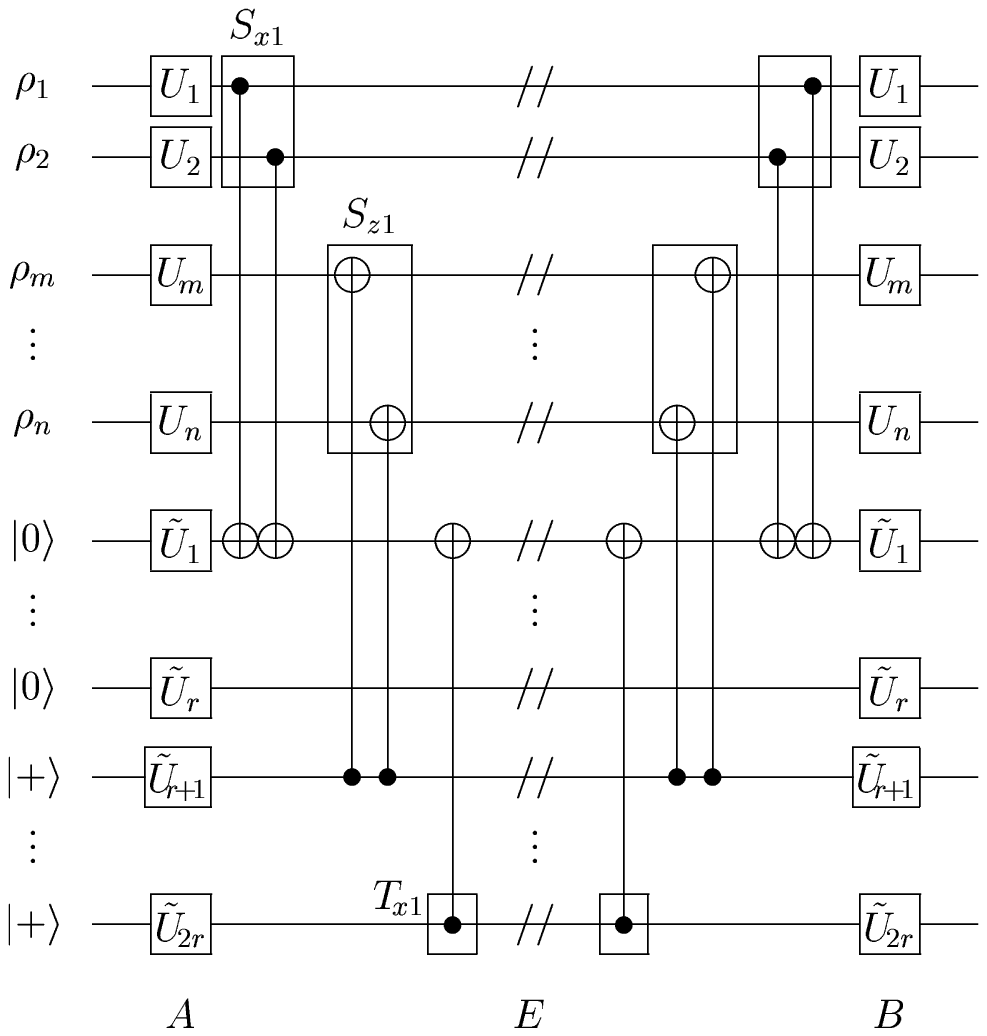,width=3.2in}}
\vspace*{3ex}
\fcaption{The modified private quantum channel.  $\rho_1,\cdots,\rho_n$
are $n$ data qubits.  Depending on the key, $U_i = I$, $X$,
$Z$, or $XZ$, 
$\tilde{U}_{1,\cdots,r} = I$ or $X$, and   
$\tilde{U}_{r+1,\cdots,2r} = I$ or $Z$.  
We only show the operations related to $S_{x1}$, $S_{z1}$, and $T_{x1}$, 
with $S_{x1}=\{1,2\}$, $S_{z1}=\{m,n\}$, and $T_{x1}=\{1\}$ as examples.
The symbol $//$ denotes a qubit in transit (and at risk).}
\label{fig:scheme3}
\end{center}
\end{figure}

\noindent
To illustrate that secure recycling is {\em not} a property special to
entanglement, we adapt a scheme in Ref.~\ncite{BBB82} to recycle the
classical key in the private quantum channel.
The main idea is to add known test qubits to detect errors effectively.
Specifically, consider sending $n$ qubits with security parameter $r$.  
Alice encodes the $n$ qubits with a $2n$-bit classical key 
as in the original scheme described in 
Section~\ref{sec:all1tp}.\ref{subsec:pqc}. 
She appends to the data qubits $2r$ test qubits, called $x_1,
\cdots, x_r, z_1, \cdots, z_r$, in the state $|0\>^{\otimes r}
|+\>^{\otimes r}$.  Each test bit may be flipped $|0\> \rightarrow
|1\>$, $|+\> \rightarrow |-\>$ depending on a $2r$-bit classical key.
In addition, she picks $2r$ random subsets
$S_{x1},\cdots,S_{xr}$, $S_{z1},\cdots,S_{zr}$ of the $n$ data qubits.
For each $i$, a {\sc cnot} is applied from each qubit in $S_{xi}$ to
$x_i$.  Likewise, a {\sc cnot} is applied from $z_i$ to each qubit in
$S_{zi}$.
Alice also picks $r$ random subsets $T_{x1}, \cdots, T_{xr}$ of $\{z_1,
\cdots, z_r\}$ and applies a {\sc cnot} from each $z_j \in T_{xi}$ to
$x_i$.\footnote{
Note that $T_{x1}, \cdots, T_{xr}$ also define $r$ random
subsets $T_{z1}, \cdots, T_{zr}$ of $x_1, \cdots, x_r$ such that 
a {\sc cnot} is applied from $z_j$ to each $x_i \in T_{zi}$.}~~  
Then, she sends all $n+2r$ qubits to Bob.
After Bob announces receipt of all the qubits, Alice announces all 
$3r$ subsets. 
Bob decodes by inverting Alice's operation.  If the test qubits are in
the state $|0\>^{\otimes n} |+\>^{\otimes n}$, he accepts the decoded
data qubits and recycles the classical key. 
The main idea behind the modification is illustrated in
Fig.~\ref{fig:scheme3}.  

If no error occurs to the $(n\!+\!2r)$-qubit cipher-text during
transmission, the test qubits are always decoded as $|0\>^{\otimes r}
|+\>^{\otimes r}$.
However, if a nontrivial Pauli error occurs, the test qubits are not 
decoded as $|0\>^{\otimes r} |+\>^{\otimes r}$ with probability 
higher than $1-2^{-r}$.
To see this, decompose the Pauli error into its $X$ and $Z$ components, 
and without loss of generality, the $X$ component is nontrivial.  
The overall effect of the extra {\sc cnot} can be found 
using the commutation relations in Section~\ref{sec:eec}.\ref{subsec:q1tpec}. 
An $X_j$ during transmission becomes $X_j$ and an extra $X_{x_i}$ on
the original cipher-text if $j \in S_{xi}$.  Likewise, $X_{z_j}$
becomes $X_{z_j}$ with an extra $X_{x_i}$ if $z_j \in T_{xi}$.
An $X_{x_i}$ decodes to itself.  
Thus $x_i$ has an overall $X$ error if an odd number of $X$ occurs to
$S_{xi} \cup T_{xi} \cup x_i$.  
As any nontrivial tensor product of $X$ errors is equally likely to
act on an even or odd number of qubits in a random subset, the
probability for $x_1$, $\cdots$, $x_r$ to decode to $|0\>^{\otimes r}$
is $2^{-r}$.
A $Z$ error is propagated to the $z_i$ similarly.  
Note that the $X$ and $Z$ components act independently on the test
qubits, and the $x_i$ are unaffected by $Z$ errors and 
the $z_i$ are unaffected by $X$ errors.  
This completes the proof that any nontrivial Pauli error is undetected 
with probability no more that $2^{-r}$. 

We now prove the security of key recycling against the most general
eavesdropping strategy.
Using the framework of Section \ref{sec:eec}.\ref{subsec:genform}, let
the received cipher-text be ${\cal E}(\rho) = \sum_{ij} e_{ij} P_i
\rho P_j^\dagger$.
Let $P_0$ be the identity Pauli error.  
Each set of random subsets corresponds to a detection scheme that 
distinguishes a set of Pauli errors ${\cal P}_I$ from its complement, 
and $P_0 \in {\cal P}_I$.  
The accepted output is ${\cal E}_a(\rho) \propto \sum_{P_i, P_j \in
{\cal P}_I} e_{ij} P_i \rho P_j^\dagger$.
Averaged over the random subsets, the unnormalized accepted state is
given by
\be
 	{\cal E}_a(\rho) = \sum_{ij} c_{ij} e_{ij} P_i \rho P_j^\dagger 
\label{eq:acceptedstate}
\ee
where $c_{ij} \leq 2^{-r}$ except for $c_{00} = 1$.  
%

The recycling scheme is secure if there is vanishing probability for
Eve to obtain a nonvanishing amount of information, $I_{\rm Eve}$, on
the recycled bits.  Since the keys are recycled only if the state is
accepted, we only need to show that the following is vanishing for 
any nonvanishing threshold $I_{\rm thres}$
\be
	{\rm Prob}\,({\rm accept~and~} I_{\rm Eve} \geq I_{\rm thres}) 
	= {\rm Prob}\,({\rm accept}) \times 
	  {\rm Prob}\,(I_{\rm Eve} \geq I_{\rm thres} | \; {\rm accept}) 
\,.
\label{eq:twoterms}
\ee	
We now show that one of the two factors in Eq.~(\ref{eq:twoterms})
has to vanish when $r$ is sufficiently large.  
Using the normalization of the accepted state in 
Eq.~(\ref{eq:acceptedstate}), 
${\rm Prob}({\rm accept}) \leq e_{00} + 2^{-r} (1-e_{00})$ can be 
made vanishing unless $e_{00}$ is nonvanishing.  
In this case, we can show that 
${\rm Prob}(I_{\rm Eve} \geq I_{\rm thres} | \; {\rm accept})$ is vanishing. 
The amount of information $I_{\rm Eve}$ is bounded by the entropy of
Eve's reduced density matrix, which in turns is bounded by the entropy of
${\cal E}_a(\rho)/{\rm tr}({\cal E}_a(\rho))$ when maximized over pure
input states $\rho$.
Rewriting the unnormalized state ${\cal E}_a(|\psi\>\<\psi|)$: 
\be
 	{\cal E}_a(|\psi\>\<\psi|) = e_{00} |\psi\>\<\psi| +  
	\sum_{(i,j) \neq (0,0)} c_{ij} e_{ij} P_i |\psi\>\<\psi| P_j^\dagger 
\,,
\ee
it can be verified that, when $e_{00}$ is nonvanishing, increasing $r$
makes the second term vanish, and 
${\cal E}_a(|\psi\>\<\psi|)/{\rm tr}({\cal
E}_a(|\psi\>\<\psi|))$ is arbitrarily close to $|\psi\>\<\psi|$ and has 
vanishing entropy.  Thus $I_{\rm Eve}$ has to vanish, and
same for ${\rm Prob}(I_{\rm Eve} \geq I_{\rm thres} | {\rm accept})$ 
for any finite $I_{\rm thres}$.

\section{The Quantum Vernam Cipher and Secret Sharing}
\label{sec:ss}

\noindent
We now explain the properties of the quantum Vernam cipher in terms of
general connections~\cite{LoGottesman00p} between secret communication
and secret sharing~\cite{Cleve99,Gottesman99b}.
A (classical or quantum) secret sharing scheme divides a secret into
{\em shares}.
The secret is retrievable only with enough shares, which form the
{\em authorized sets}.  Other sets are {\em unauthorized}.
In general, unauthorized sets can have partial information.  We
restrict to {\em perfect} schemes in which unauthorized sets have 
no information.
A $(k,n)$ {\em threshold} scheme is a perfect scheme in which 
any $k$ out of $n$ shares form an authorized set.
In addition to the usual properties, quantum secret sharing schemes
also obey the no-cloning theorem~\cite{Wootters82}, so that
complements of authorized sets are unauthorized.
Finally, in pure state perfect quantum secret sharing schemes,
complements of unauthorized sets are authorized.

Any private key encryption scheme (classical or quantum) which conveys
a message from Alice to Bob but conceals it from Eve is a secret
sharing scheme.
The secret is divided into three shares: {\bf A} and {\bf B} are
private shares for Alice and Bob, and {\bf E} is the share
communicated from Alice to Bob.  Thus {\bf A} and {\bf B} represent
the key, and {\bf E} represents the cipher-text.
By definition, $\{ {\bf A}, {\bf E}\}$ and $\{ {\bf B}, {\bf E}\}$ are
authorized while {\bf B} and {\bf E} are unauthorized.  
In a quantum cipher, {\bf A} is unauthorized.  If additionally, the
scheme is pure, $\{ {\bf A}, {\bf B}\}$ is authorized: the scheme is a
$(2,3)$ threshold scheme.

The quantum Vernam cipher is an example of pure state threshold 
scheme described above.
Entanglement is regenerated because {\bf A} and {\bf B} are 
identical shares.  Errors on {\bf E} are correctable 
because $\{ {\bf A},{\bf B} \}$ is authorized. 
Furthermore, in the quantum Vernam cipher:
({\it 1}) Alice can encode an unknown message and her half of the key
into the correctly distributed shares all by herself, and
({\it 2}) errors on {\bf E} are correctable using only local quantum
operations and classical communication (LOCC) between Alice and Bob.
We now characterize secret sharing schemes with these two properties. 
Property ({\it 1}) holds for all pure state quantum secret sharing
schemes in which the reduced density matrix of {\bf B} is maximally
mixed.
This follows from the proof of the impossibility of quantum bit
commitment~\cite{Lo97bc}, that two pure states with the same reduced
density matrix in Bob's system can be transformed to each other by
unitary operations acting outside Bob's system.
Property ({\it 2}) holds asymptotically if the entanglement between
{\bf A} and {\bf B} (in ebits) in the secret sharing scheme is at
least twice the size of {\bf E} (in qubits).  This follows from
comparing the number of errors to be distinguished with the amount of
information obtainable in the random hashing method~\cite{Bennett96a}.

\begin{figure}[ht]
\begin{center}
\mbox{\psfig{file=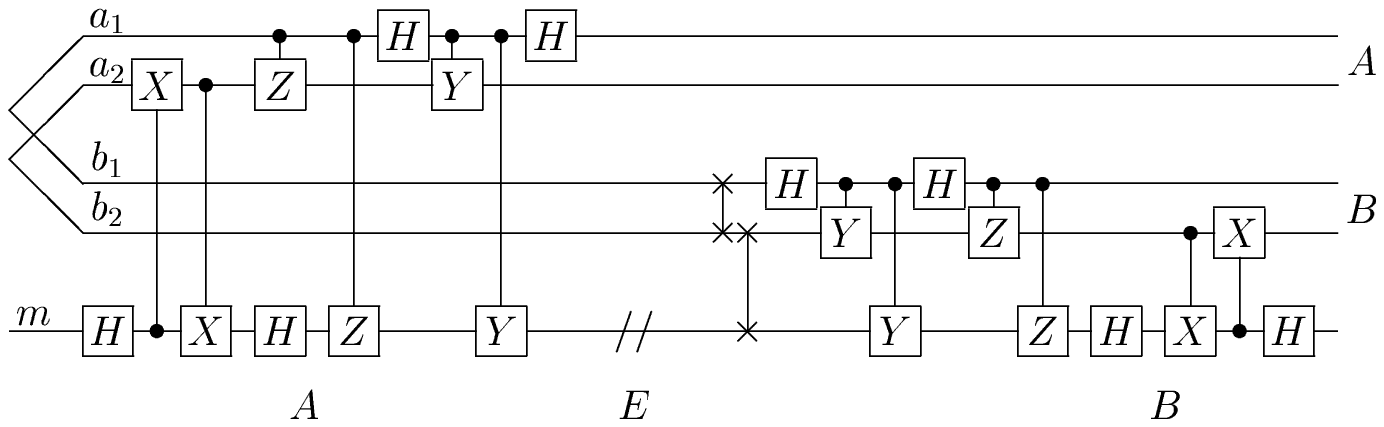,width=4.5in}}
\vspace*{3ex}
\fcaption{The $5$-bit code as a quantum cipher.  In the circuit, $Y =
iXZ$, $H = {1 \over \sqrt{2}} (X\!+\!Z)$, and a vertical line with
$\times$ in both ends is a swap operation.}
\label{fig:scheme4}
\end{center}
\end{figure}

As an example to construct a cipher from a secret sharing scheme with
the above characterization, consider the (2,3) threshold scheme
obtained from the $5$-qubit $1\mbox{-error}$ correcting
code~\cite{Bennett96a,Laflamme96}, by assigning two qubits to each of
{\bf A} and {\bf B}, and one qubit to {\bf E}.
The encoding circuit $U_{enc}$ can be specified by how the {\em
stabilizer} and the {\em encoded operations}
evolve~\cite{Gottesman97}.
As $U_{enc}$ is in the Clifford group~\cite{Gottesman97}, a circuit
implementing $U_{enc}$ can be constructed using a scheme in
Ref.~\ncite{Gottesman97}.  The decoding circuit can be constructed 
similarly.  
The cipher obtained is shown in Fig.~\ref{fig:scheme4}.

\noindent 
We find from Fig.~\ref{fig:scheme4} that the four possible Pauli
errors in the cipher-text correlate with the EPR pairs
being $|\Phi^+\>^{\otimes 2}$, $|\Phi^-\>^{\otimes 2}$,
$|\Psi^+\>^{\otimes 2}$, and $|\Psi^-\>^{\otimes 2}$, where
$|\Psi^\pm\> = {1 \over \sqrt{2}} (|01\> \pm |10\>)$.  The four cases
are distinguishable by LOCC.

Cleve~\cite{Cleve00p} derived another example of a cipher from a
secret sharing scheme (Appendix~\ref{sec:polycode}).  It is a $(2,3)$
threshold scheme in which all three shares are $3$-dimensional.
Errors on {\bf E} cannot be corrected with LOCC, unless extra
entanglement is available to Alice and Bob.  On the other hand, this
cipher requires less entanglement to conceal the message.

We can apply the connections between secret sharing and secret
communication to the private quantum channel and
teleportation~\cite{Bennett93} (see also Appendix~\ref{sec:teleport})
which encrypts a quantum plain-text to a {\em classical} cipher-text 
using a quantum key.\footnote{  
%
Teleportation paradoxically communicates quantum states {\em
securely} without quantum communication.  The closely related remote
state preparation~\cite{Lo00rsp,Bennett00} may not be secure.}~~
In teleportation, {\em after} Alice's measurements, {\bf E} is the
outcome $(k_1, k_2)$ to be communicated and {\bf B} is the quantum
state $Z^{k_2} X^{k_1}|\psi\>$ possessed by Bob.
In the private quantum channel, ${\bf A} = {\bf B} = (k_1,k_2)$ is the
classical key, and ${\bf E} = Z^{k_2} X^{k_1}|\psi\>$ is the
communicated quantum state.\footnote{
%
This provides an alternative proof for the lower bound of 
the classical key size, since an important classical share is at least 
twice the size of the quantum secret~\cite{Gottesman99b,LoGottesman00p}.}~~
Viewing {\bf A} and {\bf B} in the second scheme as one share, both
schemes are the same $(2,2)$ threshold scheme with the quantum and
classical shares interchanged.
As a {\em mixed} state $(2,2)$ scheme, errors on one share is not
correctable, as we have seen in the private quantum channel.
However, in teleportation, the classical share is broadcast and no
correction is needed.
Finally, the quantum Vernam cipher, with the three shares forming the
state ${1 \over 2} \sum_{k_1 k_2} |k_1 k_2\>~|k_1 k_2\>~Z^{k_2}
X^{k_1}|\psi\>$ is just the purification of the $(2,2)$ scheme.  

\section{Conclusion}
\label{sec:conclusion}

\noindent
We have analyzed two important properties of the quantum Vernam
cipher, the security of recycling keys and the reliability of the
transmission, and have made comparisons with other related schemes.
These results are summarized and extended to other existing schemes in
the following table~\cite{Bennett00b}, which is explained next.
%
%
\begin{center}
\begin{tabular}{l|l|l|l}
Type & Example & Security of & Reliability      
\\
$CK\!M$ & & key recycling & 
\\
\hline
CCC & Classical one-time-pad & $\times$ & $\surd$ 
\\
CCQ & Impossible &  & 
\\
CQC & Entanglement based & $\times$ & 
\\  & ~~~~~key distribution~\cite{Ekert91} & 
\\
CQQ & Teleportation & $\times$ & $\surd$  
\\[1.3ex]
QCC & Eavesdrop-detecting channel & $\surd$ & $\times$  
\\
QCQ & Private quantum channel & $\surd$ & $\times$  
\\
QQC & Superdense coding~\cite{Bennett92} & $\surd$$^*$ & $\times$  
\\
QQQ & Quantum Vernam cipher & $\surd$ & $\surd$ 
\\[1.3ex]
Q\,0\,C & BB84~\cite{BB84} &  
\\
Q\,0\,Q & Establishing entanglement &  
\end{tabular}
\end{center}
\vspace*{1ex}
\hspace*{11ex}$^*$ Requires quantum back-communication.\\[1.5ex]
%
%
\indent In the table, the type of cryptographic protocol is specified by
three elements: the communication channel (which is of the same type
as the cipher-text $C$), the key $K$, and the message $M$ to be
conveyed.  A $3$-alphabet string represents these three elements {\em
in order}.  Q, C, and 0 respectively stand for the element being
quantum, classical, and non-existing.
The first property in question is the security of key recycling and
the second property is reliability -- whether the correct message is
received with high probability.  The security properties are based on
Alice and Bob having an unjammable $2$-way classical broadcast
channel.
%

%
%
We can extrapolate the properties of the specific examples to 
classes of ciphers.  
For example, due to the use of an unjammable classical broadcast
channel, all ciphers of the type C$-$$-$ are reliable.
In contrast, ciphers of the type Q$-$$-$ are susceptible to errors,
unless a large quantum key is available, such as in the quantum Vernam
cipher.  
Since in the worst case the quantum channel is jammed, a general
recovery procedure would involve LOCC and is effectively of the type
CQQ (though the exact protocol needs not be teleportation).
The very same unjammable classical channel in C$-$$-$ cannot detect
for eavesdropping, and used key can be compromised.
The susceptibility in the quantum channel in Q$-$$-$ is also the
reason why it can detect eavesdropping and reject compromised keys.
This quantum feature also allows key distribution to be possible. 

The secure properties of the quantum Vernam cipher come at a price --
it requires a quantum channel {\em and} pre-shared entanglement.
In fact, for the same resources, one can use the quantum channel 
to establish entanglement and use the entanglement to teleport 
the state.  
The two methods are compared in Appendix~\ref{sec:resources}.
We are not aware of a circumstance in which QQQ is more efficient 
than the hybrid method Q0Q + CQQ.
This is not surprising in view of the above discussion, since the
hybrid method exploits the advantages of both types of ciphers.

We have ensured security in key recycling by detecting errors in
the cipher-text.
This objective is very similar to that of message authentication
-- to reject a forged or altered message with high probability. 
For example, our modification to the private quantum channel described
in Section \ref{sec:recycle}.\ref{subsec:ckrecycle} can be viewed as 
an authentication step for the encrypted quantum message.  
For authentication, all $U_i$ in Figure \ref{fig:scheme3} can be omitted.
The test qubits can detect both forging and tampering with high
probability due to the random flip based on the $2r$-bit classical
key.
Forging succeeds with probability no better than $2^{-r}$ and the
fidelity of an accepted message with respect to the origin cipher-text 
is of order $1 - {\cal O}(2^{-r})$.  
This means that authenticating $n$ qubits given an insecure quantum
channel and an authenticated 2-way classical channel requires only
$2r$ bits of classical key and an extra $2r$ qubits of quantum
communication.
We can also drop the assumption of authenticity in the classical
communication given a larger key to classically authenticate the
classical messages, for example, using the Wegman-Carter
method~\cite{Wegman81}.\footnote{
We need to authenticate a 1-bit message from Bob and a $(2nr+r^2)$-bit
message from Alice, requiring two keys with  
$4 (r + \log \log (2nr+r^2)) \times \log (2nr+r^2)$ and   
$4 (r + \log \log (2r+r^2)) \times \log (2r+r^2)$ bits.}~
Recently, authentication protocols for quantum message using a
classical key but no additional classical communication are
proposed~\cite{Barnum01,Crepeau01}.

Returning to the connection with secret sharing,   
we have seen that in the quantum Vernam cipher, the quantum secret can
be unlocked from an authorized set using only LOCC between the parties.  
Under the same conditions, hardly any information can be obtained 
in a recently proposed scheme~\cite{Terhal00} to share a classical
secret.
It will be interesting to understand the origin of such differences.
It might be related to the amount of entanglement shared between the
parties in the secret sharing scheme, and further investigation is
underway.
More generally, secret sharing schemes have mostly been analyzed
assuming no or full cooperation between the different parties, and
the security under LOCC remains an interesting area to be explored.

%

\nonumsection{Acknowledgements}

\noindent
Stimulating discussions, mostly during the Workshop on Quantum
Information and Computation held at the Aspen Center for Physics in
June 2000, have contributed significantly to the results presented.
The question on how to replace the classical key with a quantum one in
the private quantum channel was initially raised by Julia Kempe and
Xinlan Zhou.
We attribute various connections between secret sharing and secret
communication to interesting discussions with Hoi-Kwong Lo and Daniel
Gottesman.
We thank Hoi-Kwong Lo, Charlie Bennett, and Ike Chuang for helpful 
suggestions on recycling classical bits.  
We subsequently learned of ideas to recycle a classical key 
by Charlie Bennett, Gilles Brassard, Seth
Breidbart, and Stephen Wiesner~\cite{BBB82}, and adapted their method
in the present discussion.
We also thank Charlie Bennett for enlightening discussions on the
classification of cryptographic protocols.
We learned of quantum message authentication from Howard Barnum and 
Alain Tapp after the initial submission of the manuscript. 
We greatly appreciate informative discussions with Richard Cleve on
the cipher in Appendix~\ref{sec:polycode}, Michael Nielsen on
Shannon's classical results, John Preskill on the malleability of the
ciphers, and John Smolin on random hashing.
We thank Hoi Fung Chau, David DiVincenzo, and Michele Mosca for
enjoyable discussions, and Charlie Bennett, Hoi Fung Chau, David
DiVincenzo, Hoi-Kwong Lo, and Barbara Terhal for critical comments on
the manuscript.
We acknowledge support from the National Security Agency and the
Advanced Research and Development Activity through the Army Research
Office contract number DAAG55-98-C-0041.

%

\nonumsection{References}

%

\appendix{Definitions of some Ciphers} 
\label{sec:ciphers}

\noindent  
We briefly describe the ciphers which are not reviewed elsewhere in
this paper: 
\begin{itemlist}
\item 
{\em Entanglement based key distribution}~~Alice and Bob share a large
number of $|\Phi^+\>$.  They measure their halves of the EPR pairs
independently in the $\{|0\>,|1\>\}$ basis.  Their measurement results 
can be used as keys.  If Alice and Bob are given a quantum channel 
instead, they first establish pure entanglement with the standard 
test procedures. 
\item 
{\em BB84}~~Alice sends to Bob $\{|0\>,|1\>,|+\>,|-\>\}$ chosen at
random, and Bob measures them in random basis $\{|0\>,|1\>\}$ or
$\{|+\>,|-\>\}$.  They subsequently announce their bases.  Only the
measurement results obtained in the matching basis are used.  A
sufficient number of the results are announced and compared to test
for eavesdropping.  Upon passing the test, privacy
amplification~\cite{Bennett95} is applied to the results not announced
to establish classical keys.
\item 
{\em Superdense coding}~~Alice and Bob share one copy of $|\Phi^+\>$.
Alice can send $2$ classical bits $c_1,c_2$ securely to Bob as
follows.  Alice applies $X^{c_1}Z^{c_2}$ on her half of $|\Phi^+\>$
and sends it to Bob.  Bob can determine $c_1,c_2$ by a Bell
measurement on both qubits.
\end{itemlist}

\appendix{Recovery of Message without the Cipher-text}
\label{sec:lostm}

\noindent
Without loss of generality, let the message be $|\psi\> = a |0\> + b |1\>$.
Ordering the registers as $(a_1,b_1,a_2,b_2,m)$, the system has initial
state $|\Phi^+\>|\Phi^+\>|\psi\>$.
The state changes:
\bea 
\nonumber
	& & 
	{1 \over 2} \lbm |0000\>+|0011\>+|1100\>+|1111\> \rbm (a|0\>+b|1\>) 
\\
\nonumber
	& \rightarrow & 
	{1 \over 2} \lbm |0000\>(a|0\>+b|1\>) + |0011\> (a|0\>-b|1\>) 
\\	& & 	~	 + |1100\>(a|1\>+b|0\>) + |1111\> (-a|1\>+b|0\>) 
		    \rbm
\label{eq:eg_encode}
\\
\nonumber
	& \rightarrow &   
	{1 \over 2} \lbm a (|0000\> + |0011\>) + 
		          b (|1100\> + |1111\>) \rbm 
\\	& \oplus & 
	{1 \over 2} \lbm b (|0000\> - |0011\>) + 
			  a (|1100\> - |1111\>) \rbm 
\label{eq:eg_lostm}
\\
\nonumber
	& \rightarrow & 
	{1 \over 2} \lbm a (|0000\> + |0011\>) |0\> + 
		          b (|1100\> + |1111\>) |1\> \rbm  
\\	& \oplus &
	{1 \over 2} \lbm b (|0000\> - |0011\>) |0\> + 
			  a (|1100\> - |1111\>) |1\> \rbm 
\label{eq:eg_insertm}
\\
\nonumber
	& = & 
	{1 \over 4} \lbm (|00\> + |11\>)(|00\> + |11\>) (a|0\> + b|1\>) 
\\	& & ~~ +         (|00\> - |11\>)(|00\> + |11\>) (a|0\> - b|1\>) 
\nonumber
		    \rbm  
\\	& \oplus &  
	{1 \over 4} \lbm (|00\>+|11\>)(|00\>-|11\>) (b|0\>+a|1\>) 
\nonumber
\\	& & ~~ +   	 (|00\>-|11\>)(|00\>-|11\>) (b|0\>-a|1\>) \rbm 
\label{eq:eg_decode}
\eea
describe the encoding (Eq.~(\ref{eq:eg_encode})), the removal of $m$ 
(Eq.~(\ref{eq:eg_lostm})), and the decoding by Bob after he substitutes  
$|0\>$ for $m$ (Eq.~(\ref{eq:eg_insertm})).  
The $\oplus$ denotes a mixture of states: $\oplus_i |\psi_i\> \equiv 
|\psi_i\>\<\psi_i|$.  
The decoded state is rewritten in Eq.~(\ref{eq:eg_decode}), to which
the syndrome measurement described in
Section~\ref{sec:eec}.\ref{subsec:q1tpec} is applicable.

\appendix{Quantum Secret Sharing Scheme as Secure Quantum Channel}
\label{sec:polycode}

\noindent
We describe another cipher due to Cleve~\cite{Cleve00p} constructed
{from} a $(2,3)$ threshold quantum secret sharing scheme.
The plain-text $|\psi\> = \alpha |0\> + \beta |1\> + \gamma |2\>$ is
a three dimensional state (a qutrit).
We define the following gates acting on qutrits: 
\bea
\mbox{\psfig{file=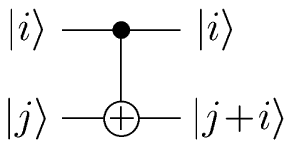,width=0.9in}
    ~~\psfig{file=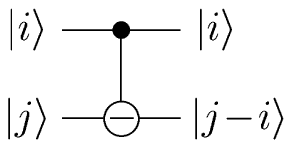,width=0.9in}
    ~~\psfig{file=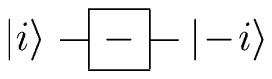,width=0.9in}}	
\nonumber
\eea
where sums and differences are taken modulo $3$.  
The proposed scheme can be represented by the following circuit: 
\bea
\mbox{\psfig{file=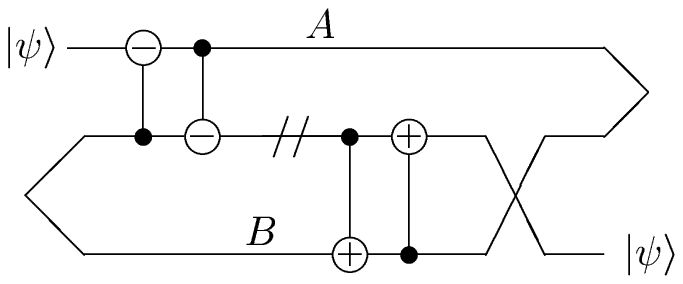,width=2.6in}}
\nonumber
\eea
in which the maximally entangled state is ${1 \over \sqrt{3}} (|00\> +
|12\> + |21\>)$, $A$, $B$ represent the private shares of Alice and
Bob, and $//$ represents a transmission from Alice to Bob.
The regenerated entangled state is explicitly marked.
Encoding is performed locally by Alice.
As a $(2,3)$ threshold scheme, any error in the transmitted qutrit is
correctable.  However, correction {\em cannot} be performed using only
LOCC operations by Alice and Bob.
To see this, we first rearrange the qutrits in the circuit and redefine 
the maximally entangled state as ${1 \over \sqrt{3}} (|00\> +
|11\> + |22\>)$.
\bea
\mbox{\psfig{file=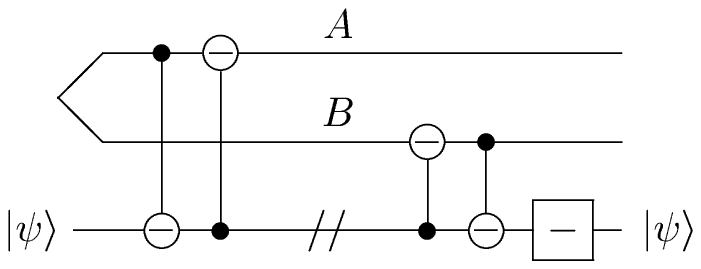,width=2.6in}}
\nonumber
\eea
We can now easily find the effect of an error $\cal E$ during
transmission, for the following circuits are equivalent:
\bea
\centerline{\mbox{\psfig{file=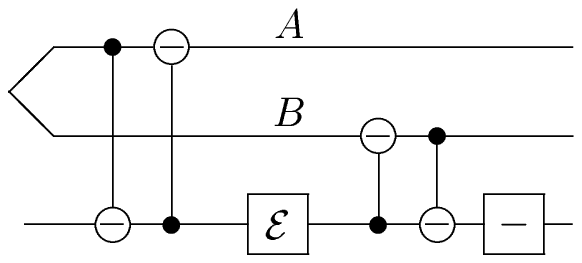,width=2.0in}}}
\nonumber
\\
\nonumber
\\
\centerline{\mbox{\psfig{file=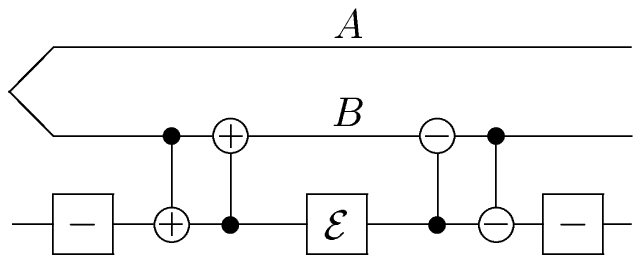,width=2.1in}}}
\nonumber
\eea
We consider an error basis on a qutrit generated by $X$ and $Z$ where 
$X|j\> = |j\!+\!1\>$ and $Z|j\> = e^{2 \pi i j/3} |j\>$.   
Using the commutation relations 
\bea
\mbox{\psfig{file=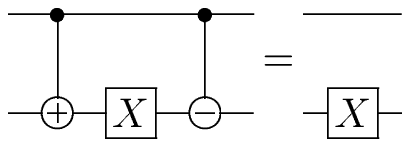,width=1.3in}}\,,~~~~~~~~
\mbox{\psfig{file=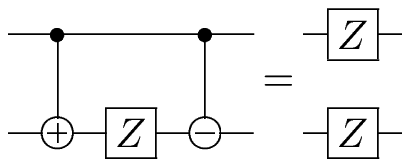,width=1.3in}}\,,
\nonumber
\\[0.2ex]
\nonumber
\\
\mbox{\psfig{file=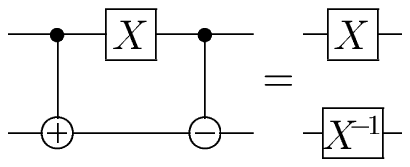,width=1.3in}}\,,~~~~~~~~
\mbox{\psfig{file=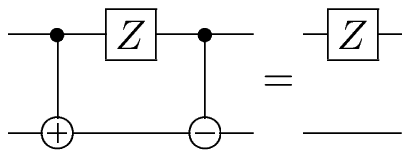,width=1.3in}}\,,
\nonumber
\vspace*{1ex}
\eea
the overall effects due to the errors $X^t$ and $Z^t$ can be obtained: 
\bea
\mbox{\psfig{file=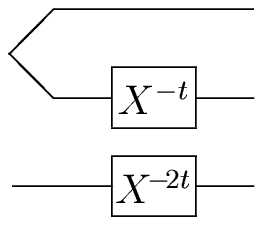,width=0.7in}}
\hspace*{2ex}\begin{array}{c}{,}\\{~}\\{~}\end{array}
\hspace*{10ex}\mbox{\psfig{file=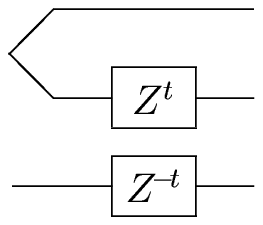,width=0.7in}} 
\nonumber
\eea
The $9$ possible errors are correlated with $9$ orthogonal maximally
entangled states, which are globally distinguishable but
indistinguishable with LOCC, or else Alice and Bob can identify
maximally entangled states from the maximally mixed state and distill
entanglement out of nothing.

\appendix{Teleportation}
\label{sec:teleport}

\noindent
Without loss of generality, consider the teleportation of a pure state 
$|\psi\> = a |0\> + b |1\>$ using the following circuit: 
\bea
\nonumber
\centerline{\mbox{\psfig{file=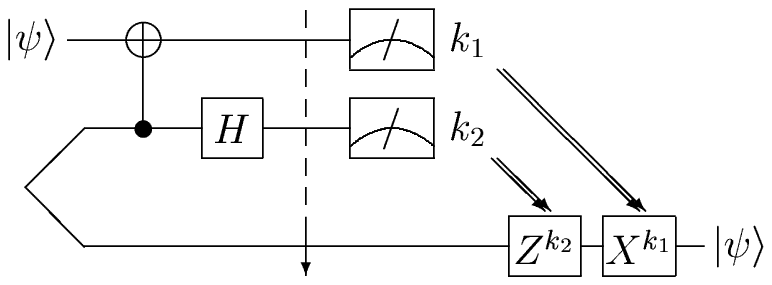,width=2.6in}}}
\eea
It is easily verified that the initial state 
${1 \over \sqrt{2}} (a|0\> + b|1\>)(|00\>+|11\>)$ is transformed to 
\bea
	& & {1 \over 2} \lbm 
	|00\>(a|0\>+b|1\>) + |01\>(a|0\>-b|1\>) 
\\      & & ~~ + |10\>(a|1\>+b|0\>) + |11\>(-a|1\>+b|0\>) \rbm 
\\
	& = & {1 \over 2} \sum_{k_1 k_2} |k_1\>|k_2\> Z^{k_2} X^{k_1} |\psi\>
\eea
right before measurement.  
The measurement results $k_1,k_2$ are sent over a classical channel to 
recover $|\psi\>$.  

\appendix{Comparison of Resources} 
\label{sec:resources}

\noindent
We compare the asymptotic resources required to send $n$ qubits
securely by (1) the quantum Vernam cipher (QQQ) and (2) 
establishing entanglement and teleporting (Q0Q + CQQ).
We compare the net amount of entanglement consumed, allowing both
schemes $n$ uses of an insecure quantum channel and unlimited uses of
a 2-way classical broadcast channel.
The quantum Vernam cipher uses $2n(1-F)$ ebits where $F$ is the
recyclable fraction of entanglement.  Teleportation uses $n(1-D_2)$
ebits where $n D_2$ ebits are distillable from $n$ uses of the quantum
channel.
Hence, teleportation is more efficient if and only if 
$F \leq (1+D_2)/2$.
%


In the following comparisons, we use more optimal recycling 
strategies than that in Section~\ref{sec:recycle}.\ref{subsec:qkrecycle}.  
Without eavesdropping, $F \approx D_2 \approx 1$.
If Eve measures every qubit in the computation basis, $Z$ occurs
randomly.  Hence $D_2 = 0$ and $F = 1/2$ since the EPR pairs 
detecting $X$ errors are intact.
If $I$, $X$, $Z$, $XZ$ occur with probabilities $1/2$, $1/6$, $1/6$,
$1/6$, $D_2 = 0$ and $F = 0.1037$.
For a completely random Pauli channel, $D_2 = F = 0$.  
Hence for the first two cases, the two methods are equally efficient. 
For the last two cases, teleportation is much more efficient. 

%


\begin{thebibliography}{000}
\bibitem{Note} The quantum Vernam cipher randomizes 
	the quantum plain-text in two conjugate bases.  The scheme in
	Ref.~\ncite{Zhang00} randomizes the quantum plain-text in one
	basis, and does not conceal it completely.
\bibitem{interfere} In the noise process: $|\psi\>\<\psi| \rightarrow
\sum_k D_k |\psi\>\<\psi| D_k^\dag$, the $D_k$ are said to be
noninterfering since the terms $D_k |\psi\>$ do not exhibit
interference.  In contrary, in the process: $|\psi\>\<\psi|
\rightarrow \sum_{ij} e_{ij} P_i |\psi\>\<\psi| P_j^\dag$, the 
terms $P_i |\psi\>$ interfere with each other, and $P_i$ are 
said to be interfering errors.
\end{thebibliography}
\end{document}